\documentclass{article}
\usepackage{emulateapj}
\usepackage{apjfonts}
\usepackage{graphics}

\newenvironment{inlinefigure}{%
\def\@captype{figure}%
\noindent\begin{minipage}{0.999\linewidth}\begin{center}}
{\end{center}\end{minipage}}

\newlength{\colwidth}
\newcommand{\cm}{\rm cm} 
\newcommand{\s}{{\rm s}}
\newcommand{\kms}{{\rm km}\,{\rm s}^{-1}}
\newcommand{\K}{{\rm K}}
\newcommand{\kpc}{{\rm kpc}}
\newcommand{\Mpc}{{\rm Mpc}}
\newcommand{\erg}{{\rm erg}}
\newcommand{\Msun}{{{\rm M}_\odot}}

\newcommand{\HeII}{\ion{He}{2}}

\newcommand{\lya}{Ly$\alpha$} 

\lefthead{Schaye}
\righthead{The nature of the Ly$\alpha$ forest} 

\begin{document}

\submitted{Accepted for publication in the Astrophysical Journal}

\title{Model-independent insights into the nature of the
Ly$\alpha$ forest and \\ the distribution of matter in the universe} 
\author{Joop~Schaye}
\affil{School of Natural Sciences, Institute for Advanced
Study, Einstein Drive, Princeton NJ 08540, schaye@ias.edu}

\begin{abstract}
Straightforward physical arguments are used to derive the properties
of \lya\ forest absorbers. It is shown that many aspects of the
current physical picture of the forest, in particular the
fact that the absorption arises in extended structures of moderate
overdensities which contain a large fraction of the baryons in the
universe, can be derived directly from the observations without making
any specific assumptions about the presence and distribution of dark
matter, the values of the cosmological parameters or the mechanism for
structure formation.  The key argument is that along any line of sight
intersecting a gravitationally confined gas cloud, the size of the
region over which the density is of order the maximum density, is
typically of order the local Jeans length. This is true for overdense
absorbers, regardless of the shape of the cloud and regardless of
whether the cloud as a whole is in dynamical equilibrium.  The simple
analytic model is used to derive the mass distribution of the
photoionized gas directly from the observed column density
distribution. It is demonstrated that the shape of the column density
distribution, in particular the observed deviations from a single
power-law, and its evolution with redshift, reflect the shape of the
matter distribution and can be understood in terms of the growth of
structure via gravitational instability in an expanding universe.
\end{abstract}

\keywords{cosmology: theory --- galaxies: formation ---
intergalactic medium --- quasars: absorption lines --- hydrodynamics}

\section{Introduction}

The \lya\ forest is thought to arise in the smoothly fluctuating
intergalactic medium (IGM), which traces the distribution of the dark
matter on large scales. Mildly overdense regions give rise to low
column density absorption lines ($N_{HI} \la 10^{14.5}\,\cm^{-2}$),
whose widths are determined mainly by thermal broadening and the
differential Hubble flow across the absorber. Although many aspects of
this physical picture of the \lya\ forest can already be found in
papers by Bi, B\"orner, \& Chu (1992) and Bi (1993), who used a
semi-analytic model with a log-normal distribution for the dark
matter; it gained acceptance only after it was confirmed and extended
by hydrodynamical simulations (e.g., Cen et al.\ 1994; Zhang, Anninos,
\& Norman 1995; Hernquist et al.\ 1996; Miralda-Escud\'e et al.\ 1996;
Theuns et al.\ 1998). In the simulations the \lya\ forest arises in a
network of sheets, filaments, and halos, which give rise to absorption
lines of progressively higher column densities.
This physical picture of the forest is supported by
observations of gravitationally lensed quasars and quasar pairs, which
show that the characteristic sizes of the absorbers are 10s to 100s of
kpc (e.g., Bechtold et al.\ 1994; Dinshaw et al.\ 1994, 1995; Smette
et al.\ 1995) and by the success of cosmological simulations in
reproducing the statistics of the observed absorption spectra (see
Efstathiou, Schaye, \& Theuns 2000 for a recent review).

Following Rees (1986), most theoretical work on the \lya\ forest has
been carried out within the framework of the cold dark matter
paradigm. In virtually every case an ab-initio approach was followed,
i.e.\ the properties of the forest were predicted for a particular
model of structure formation and then compared to observations (but
see for example Weinberg et al.\ 1997 and Nusser \& Haehnelt 1999 for
notable exceptions). The models, which can be either semi-analytic or fully
numerical, are specified by the cosmological parameters, the initial
power spectrum, and the evolution of the UV background radiation. 

In this paper a rather different approach will be taken. It will be
shown that many properties of the absorbers, including their densities
and radial sizes, can be derived using straightforward physical
arguments, independent of the details of the cosmological model.  The
key physical argument is that the absorbers will generally not be far
from local hydrostatic equilibrium, i.e.\ the size of the region over
which the density is of order the maximum density along the sightline,
is of order the local Jeans length. It will be argued that this is
very likely the case for overdense absorbers, regardless of whether
the cloud as a whole is in hydrostatic equilibrium.  The same
formalism will be used to explain the observed structure in the column
density distribution and to derive the density distribution of the
photoionized IGM directly from the observations.

The main physical arguments underlying this work are presented in
section \ref{sec:hydrodynamics}. Although the expressions derived in
section \ref{sec:hydrodynamics} apply to all gravitationally confined
gas clouds, only photoionized, optically thin
clouds ($N_{HI} \la 
10^{17}\,\cm^{-2}$) are discussed in 
subsequent sections. The properties of self-shielded clouds, which are
relevant for protogalactic clumps, Lyman limit systems, and damped
\lya\ absorbers, are derived in Schaye (2001). The formalism of
section \ref{sec:hydrodynamics} is used to derive the physical
properties of \lya\ forest absorbers in section
\ref{sec:properties}. The cosmological implications of the observed column
density distribution are discussed in section
\ref{sec:implications}. Finally, the main conclusions are summarized
in section \ref{sec:conclusions}. Throughout the paper, numerical
values will be inserted for physical parameters whose values are
thought to be well known. The full expressions can be found in the
appendix.


\section{Hydrodynamics}
\label{sec:hydrodynamics}
In this section the physical properties of self-gravitating gas clouds
will be derived. It will be argued that along each sightline through
an overdense absorber, the sound crossing timescale $t_{\rm sc}$, is
of order the local dynamical timescale $t_{\rm dyn}$. In other words,
the characteristic size of the absorber is typically of order the
local Jeans length. The characteristic size is defined as the size of
the region for which the density is of order the characteristic
density of the absorber. In section 
\ref{sec:chardens} it will be shown that because most of the
absorption takes place in the densest gas along the sightline, a
well-defined characteristic density does exist.

Consider a cloud with characteristic density $n_H$. 
The dynamical (or free-fall) time is
\begin{equation}
t_{\rm dyn} \equiv {1 \over \sqrt{G\rho}} 
\sim 1.0 \times 10^{15} ~\s~ \left ( {n_H \over 1 ~\cm^{-3}} \right )^{-1/2}
\left ( {1-Y \over 0.76} \right )^{1/2}
\left ( {f_g \over 0.16} \right )^{1/2},
\label{eq:tdyn}
\end{equation} 
where $f_g$ is fraction of the mass in gas\footnote{Stars and
molecules do \emph{not} contribute to $f_g$.} and
$Y$ is the baryonic mass fraction in helium, which will be set equal
to 0.24 hereafter. In cold, collapsed clumps $f_g$ could be close to
unity, but on the scales of interest here $f_g$ will not be far
from its universal value, $f_g \approx \Omega_b / \Omega_m$. 
Let $L$ be the characteristic size of the cloud,
i.e.\ the length of the intersected part of the cloud over which the 
density is of order the characteristic density.
The sound crossing time is then
\begin{equation}
t_{\rm sc} \equiv {L \over c_s} \sim 2.0 \times 10^{15}~\s~
\left ({L \over 1~\kpc} \right ) T_4^{-1/2} \left (\mu \over
0.59\right )^{1/2},
\label{eq:tsc}
\end{equation} 
where $c_s$ is the sound speed in an ideal, monatomic gas with the
ratio of specific heats $\gamma = 5/3$, $T \equiv 
T_4 \times 10^4~\K$, and $\mu$ is the mean molecular weight. In what
follows $\mu$ will be set equal to the
value appropriate for a fully ionized, primordial plasma, $\mu =
4/(8-5Y) \approx 0.59$.

The equation for hydrostatic equilibrium is $dP/dr = - G\rho
M/r^2$, where $M$ is the mass interior to $r$. Since the pressure $P
\sim c_s^2 \rho$, this 
implies $c_s^2 \rho / L \sim G\rho^2 L$, i.e.\ $t_{\rm sc} \sim t_{\rm
dyn}$. There are many other ways of showing that in
hydrostatic equilibrium $t_{\rm sc} \sim t_{\rm dyn}$. It can for
example readily be seen that it implies that the sound speed is of
order the circular velocity.  
The condition $t_{\rm sc} = t_{\rm dyn}$ defines a characteristic
length, the so-called Jeans length\footnote{The Jeans length or,
equivalently, the dynamical timescale is not well defined and some
authors include a dimensionless coefficient of order unity in the
definition. However, the value of the dimensionless coefficient
depends on the geometry of the cloud and it is hard to justify any
particular choice.},
\begin{equation} 
L_J \equiv {c_s \over \sqrt{G\rho}}
\sim 0.52 ~\kpc ~ 
n_H^{-1/2} T_4^{1/2}
\left ( {f_g \over 0.16} \right )^{1/2}.
\label{eq:LJ}
\end{equation} 
To make a connection with observations, it is useful to
introduce the `Jeans column density',
\begin{equation}
N_{H,J} \equiv n_H L_J \sim 1.6 \times 10^{21} ~\cm^{-2}~
n_H^{1/2} T_4^{1/2}
\left ( {f_g \over 0.16} \right )^{1/2},
\label{eq:NJ}
\end{equation} 
If $t_{\rm sc} \gg t_{\rm dyn}$, then the cloud is Jeans unstable and
will either fragment or, since $v 
\sim L / t_{\rm dyn} \gg c_s$, shock to the virial
temperature. In either case, equilibrium will be restored on the 
dynamical timescale.
If, on the other hand, $t_{\rm sc} \ll t_{\rm dyn}$, then the cloud
will expand or evaporate and equilibrium will be restored on the sound
crossing timescale. 

Although the Jeans criterion for gravitational instability is
familiar, it is not always appreciated that a growing density
perturbation will generally be close to `local hydrostatic
equilibrium', i.e.\ $t_{\rm sc} \sim t_{\rm dyn}$ locally, but not
necessarily for the cloud as a whole. In other words, along any
sightline through the evolving cloud, which will in general not be
spherical, the length of the region for which the density is of order
the maximum density, will be of order the local Jeans length. If the
cloud has substructure, then the condition of local hydrostatic
equilibrium should be applied to each density maximum along the
sightline. Note that the concept of a Jeans mass only makes sense if
the cloud is roughly spherical. The collapse will in general not be
synchronized along all three spatial dimensions, and a density
perturbation may first collapse into a sheet-like or filamentary
structure before reaching a more spheroidal
configuration. Furthermore, the tidal field around collapsed regions
can induce the formation of a network of sheets and filaments. Although 
the mass of the region containing the perturbation can thus differ
substantially from the Jeans mass,
\emph{along any sightline intersecting an evolving perturbation, the
characteristic size will generally be of order the local Jeans
length}.

Large departures from local hydrostatic equilibrium do occur when the pressure
changes on a timescale much shorter than $t_{\rm dyn}$. For example,
during shell crossing the density suddenly increases by a large
factor. This will result in shocks, leading to virialization and
equilibrium will quickly be restored. A sudden drop in the temperature
through a thermal instability is another example. In that case
equilibrium will be restored through fragmentation. These two examples
are relevant for high column density absorbers and are discussed
in more detail in Schaye (2001). For the \lya\ forest, sudden heating
is in fact more relevant. During reionization the temperature of the
diffuse IGM is suddenly raised by several orders of magnitude to at
least $10^4\,\K$ and possibly $\ga 10^5\,\K$, depending on the speed
of reionization and the spectrum of the ionizing radiation. Because of
the sudden increase in the Jeans length during reionization, $t_{\rm
sc} \ll t_{\rm dyn}$ for most of the gas. Until pressure forces 
have had time to restore local hydrostatic equilibrium, the gas can be
clumpy on 
scales smaller than the Jeans length (e.g., Gnedin \& Hui
1998). Simulations by Gnedin (2000) show that even if the IGM was
reionized as late as $z\sim 7$, the baryon smoothing scale is again
similar to the Jeans scale by the redshifts of interest here ($z \le
4$).

Pressure forces can only smooth the gas distribution on scales smaller than
the sound horizon, which is defined by the condition $t_{\rm sc} \sim
H^{-1}$. Hydrostatic equilibrium would then require $t_{\rm dyn}
\sim H^{-1}$, which implies $\rho \sim \bar{\rho}$. Hence, the
assumption of local hydrostatic equilibrium may break down for
absorbers with characteristic densities smaller than the cosmic mean. 
The precise value of the sound horizon depends on the thermal
history of the IGM. After reionization, the temperature of the
low-density IGM is expected to drop. At very high redshift ($z \ga 7$)
the IGM can cool efficiently through inverse Compton scattering off the
microwave background, but after that the main cooling mechanism is adiabatic
expansion and the cooling time is thus of order the Hubble time. 
Hence, for $z < z_{\rm reion}$, the sound horizon is greater than
$c_s/H$, which implies that the statement that local hydrostatic
equilibrium is only a good approximation for overdense absorbers, is
conservative.

Finally, if a cloud is not gravitationally supported, then local
hydrostatic equilibrium does not necessarily imply that its
characteristic size is of order the local Jeans length. A cloud that
is not supported by self-gravity can be pressure confined,
rotationally supported or it can be not supported at all. 
Although pressure confinement may be important for the high column
density absorption lines, the
possibility that a significant fraction of the \lya\ forest arises in
pressure confined clouds has been ruled out on both observational and
theoretical grounds (e.g., Rauch 1998). Even if every \lya\ absorption
line were to arise in a rotationally supported cloud, then the
characteristic size will only be much greater than the Jeans length
for sightlines that are nearly perpendicular to the spin axis of the
cloud, which cannot be true for a large fraction of the absorption
lines.  As discussed above, unless the cloud is kept in a steady state
by external pressure or by a centrifugal barrier, local hydrostatic
equilibrium is restored on a timescale $t \ll H^{-1}$ if the absorber
is overdense.  Hence, if a population of \lya\ forest
absorbers exists that are quasi-stable, i.e.\ whose lifetimes are not
much shorter than the Hubble time, then any given absorption line is
unlikely to arise in a 
transient cloud, i.e.\ a cloud with a lifetime $t \ll H^{-1}$,
unless most of the matter in the universe is
contained in clouds far from dynamical equilibrium.  In summary, it is
very likely that local hydrostatic equilibrium is a good approximation
for most \lya\ forest absorbers.

\subsection{The characteristic density}
\label{sec:chardens}

The characteristic density is the column density weighted density of
an absorber.
It is easily demonstrated that for a wide range of density profiles, the
neutral hydrogen column density through a cloud is dominated by the
region with the highest density. For a spherical cloud with a density
profile $n_H \propto r^{-n}$, the neutral hydrogen column density in
a sightline with impact parameter $b$ is
\begin{equation}
N_{HI} \propto \int_{-\infty}^{+\infty} (b^2 + l^2)^{-n}\,l\,d\ln l,
\end{equation}
where I used the fact that the neutral fraction in an
optically thin gas is proportional to the density.
For $n>1/2$, the dominant contribution to the column density
comes from $l \la b$ and $N_{HI} \sim n_{HI}(b)\,b$. For example, for
a singular isothermal sphere ($n=2$), the column density weighted
density is  
$
\left < n_H \right > \equiv \int n_H n_{HI}\,dl / \int n_{HI}\,dl 
= {3 \over 4} n_H(b).
$
Hence, for reasonable density profiles, almost all the absorption
takes place in the densest part of the gas cloud. This
implies the existence of a characteristic density, which will be close
to the maximum density along the line of sight through the cloud.

\section{Physical properties}
\label{sec:properties}
In the previous section expressions were derived for the size and total
hydrogen column density of an absorber as a function of its density
and temperature. To make contact with observations it is necessary to
compute the neutral fraction since it is the neutral hydrogen column density
that is observed. Here, only optically thin clouds
($N_{HI} \la 10^{17}~\cm^{-2}$) will be considered, for which
the ionization correction can be computed analytically.

The neutral fraction of a highly ionized, optically thin gas is
\begin{equation}
{n_{HI} \over n_H} = n_e \beta_{HII} \Gamma^{-1} 
\sim  0.46 ~ n_H T_4^{-0.76} \Gamma_{12}^{-1}, 
\label{eq:neutralfrac}
\end{equation}
where $\beta_{HII} \approx 4 \times 10^{-13} T_4^{-0.76}
~\cm^3\,\s^{-1}$ and $\Gamma \equiv \Gamma_{12} \times
10^{-12}~\s^{-1}$ are the hydrogen  
recombination and photoionization rates respectively and $n_e$ is the
number density of free electrons. The photoionization rate
corresponding to a UV background intensity of $J(\nu)
~\erg\,\s^{-1}\,\cm^{-2}\,{\rm sr}^{-1}\,{\rm Hz}^{-1}$ is
\begin{equation}
\Gamma = \int_{\nu_L}^\infty {4\pi J(\nu)\sigma(\nu) \over h\nu}\,d\nu,
\end{equation}
where $\sigma$ is the cross section for photoionization and $\nu_L$ is
the frequency at the Lyman limit. At redshifts $2 \la z \la 4$,
studies of the  
proximity effect find $\Gamma \sim 10^{-12}~\s^{-1}$ with a relative
uncertainty of order unity (Scott et al.\
2000 and references therein), which is in reasonable agreement with
models of the UV background from quasars (e.g., Haardt \& Madau 1996).  
At low redshift ($z \la 0.5$) the intensity of the UV background is
thought to be considerably lower, $\Gamma \approx
10^{-14}\,$--$\,10^{-13}~\s^{-1}$ (e.g., Dav\'e \& Tripp 2001 and
references therein). 

To make further progress, an estimate of the temperature is
needed. It has been known since long that the widths of
\lya\ absorption lines are consistent with photoionization
temperatures, $T\sim 10^4~\K$ (e.g., Carswell et al.\ 1984).
Gas with a temperature significantly lower
than this is heated on the photoionization timescale, $t = 
\Gamma^{-1} \ll H^{-1}$. To reach temperatures $T \gg
10^4~\K$, the gas would have to be shock-heated. Accretion shocks
associated with gravitational collapse are likely to be important only
for gas at high overdensities, but galactic winds could in principle
heat a large fraction of the IGM. For densities $n_H \ga
10^{-4}\,\cm^{-3}$ ($\delta \ga 10$ at $z\sim 3$), atomic 
cooling is efficient\footnote{The cooling time depends on the neutral
fraction and thus on $\Gamma$. If $\Gamma_{12} \ll 1$, as is the case
at low redshift, then atomic cooling will become efficient at much
lower densities than $10^{-4}\,\cm^{-3}$.} and the temperature will be
approximately 
$10^4\,\K$ (molecular cooling can lead to lower temperatures, but the
molecular fraction is negligible for optically thin clouds).
Cooling can become inefficient if the temperature is very high, $T
\gg 10^5 ~\K$, but in that case collisional ionization suppresses
the neutral fraction and the gas will only give rise to lines of very
low column density. In short, $T\sim 10^4$ is likely to be a good
approximation for the \lya\ forest, although a somewhat higher
temperature could be appropriate at $n_H \la 10^{-4}\,\cm^{-3}$ if
galactic winds are important.

\subsection{Densities}
\label{sec:densities}

Substituting equation \ref{eq:neutralfrac} into equation \ref{eq:NJ}
yields a relation between the neutral hydrogen column density and the
characteristic density, 
\begin{equation}
N_{HI} \sim 2.3 \times 10^{13} ~\cm^{-2}~ 
\left ({n_H \over 10^{-5}~\cm^{-3}}\right )^{3/2} T_4^{-0.26} \Gamma_{12}^{-1}
\left ({f_g \over 0.16}\right )^{1/2}.
\label{eq:NHIa}
\end{equation}
Because the gas responsible for the low-column density \lya\ forest is
still expanding, it is often instructive to use the density
contrast, $\delta \equiv (n_H - \bar{n}_H)/\bar{n}_H$, where
\begin{equation}
\bar{n}_{\rm H} \approx 1.1 \times 10^{-5}~ \cm^{-3}~
\left ({1+z \over 4}\right)^3
\left ( {\Omega_b h^2 \over 0.02} \right ),
\label{eq:n_H} 
\end{equation}
instead of $n_H$,
\begin{eqnarray}
N_{HI} &\sim & 2.7 \times 10^{13} ~\cm^{-2} ~ (1+\delta)^{3/2}
T_4^{-0.26} \Gamma_{12}^{-1} \nonumber \\
&& \left ({1+z \over 4}\right )^{9/2}
\left ({\Omega_b h^2 \over 0.02}\right )^{3/2}
\left ({f_g \over 0.16} \right )^{1/2}.
\label{eq:NHIb}
\end{eqnarray}
Clearly, the low-column density \lya\ forest arises in gas of low
overdensity. Virialized minihalos have $\delta \ga 10^2$ which
corresponds to high column density absorption lines, $N_{HI} >
10^{16}~\cm^{-2}$ at $z\sim 3$. The presence of dark matter does not
have a dramatic effect on the inferred densities: if $f_g = 1$ instead
of 0.16 ($\approx \Omega_b/\Omega_m$), then
the density contrast corresponding to a fixed column density is only
about a factor 2 smaller.  

The neutral hydrogen column density
corresponding to a fixed density contrast is about a factor
$5\times 10^2\,\Gamma(z=0)/\Gamma(z=3) \sim 5$--50 lower at $z\sim 0$
than at $z\sim 3$, with only a weak dependence on the evolution of the
temperature. Because the strong dependence on
redshift is partially offset by the evolution of $\Gamma$, the local
\lya\ forest is not fundamentally different from the $z\sim 3$ forest.

Hydrodynamical simulations have been used to make detailed predictions
for the low column density \lya\ forest. Although the formalism
developed here does not make any specific assumptions about the
cosmology or the nature of the dark matter, it can of course be
applied to the specific models tested with the simulations. 
In the simulations (which generally do not include feedback from star
formation or radiative transfer effects) it is found that the
low-density gas ($n_H \la 
10^{-4}~\cm^{-3}, \delta \la 10$ at $z\sim 3$) follows a tight
temperature-density relation, which can be approximated by a
power-law, $T=T_0(1+\delta)^\alpha$ (Hui \& Gnedin 1997). 
Substituting this relation into equation \ref{eq:NHIb} yields,
\begin{eqnarray}
N_{HI} &\sim & 2.7 \times 10^{13} ~\cm^{-2} ~ (1+\delta)^{1.5-0.26\alpha}
T_{0,4}^{-0.26} \Gamma_{12}^{-1} \nonumber \\
&& \left ({1+z \over 4}\right )^{9/2}
\left ({\Omega_b h^2 \over 0.02}\right )^{3/2}
\left ({f_g \over 0.16} \right )^{1/2} \label{eq:NHIc} \\
&& (n_H \la 10^{-4}~\cm^{-3}) \nonumber.
\end{eqnarray}
From comparisons of the line widths in simulated and observed spectra
it is found that $T_0 \sim 2\times 10^4\,\K$ and $\alpha \sim
0.0$--0.3 at $z \sim 3$ (e.g., Schaye et al.\ 2000; McDonald et al.\
2001). 

Dav\'e et al.\ (1999, hereafter DHKW) find that the following empirical
relation fits the results of their simulation over the range $z = 0$--3:
\(
N_{HI} \sim 10^{14}~\cm^{-2}~ \left [{1 \over 20} (1+\delta) 10^{0.4z}
\right ]^{1/0.7},
\)
although the scatter is large for $z \la 1$.
This relation is of course specific to the simulation of DHKW, i.e.\ it
would have been different if they had used a simulation with a
different cosmology, UV background or temperature. Nevertheless, we
can compare it to equation \ref{eq:NHIc} if we set all the other
parameters to the values of DHKW's simulation\footnote{The simulation
of DHKW has $\Omega_b h^2 = 0.02$, $\Omega_b/\Omega_m = 0.12$ ($=
f_b$), $(T_{0,4},\alpha) \approx (1.0,0.5)$ at $z=3$. They use the
model of Haardt \& Madau (1996) for the evolution of the ionizing
background, but multiply the ionization rate by a factor 1.34.}.
The normalization ($N_{HI}(\delta = 0)$) of DHKW turns out to be about a
factor 3 higher ($7.2 \times 10^{13}$ vs.\ $2.3 \times 10^{13}$),
which is about as good as could be expected 
considering that there is an uncertainty of order unity in 
the normalization of both equation \ref{eq:NHIc} (because of some
arbitrariness in the value of the dimensionless coefficient in the
definition of the Jeans length) and the empirical relation of DHKW (because 
of the arbitrariness in the definition of the density contrast
corresponding to an absorption line in a simulated spectrum). 
In fact, the normalizations found by different groups vary by about an
order of magnitude, with most groups finding somewhat
lower values for $N_{HI}(\delta = 0)$ than DHKW (e.g., Bryan et al.\
1999; Schaye et al.\ 1999). Equation \ref{eq:NHIc} also reproduces the
scaling of $N_{HI}$ with $\delta$ and $z$ remarkably well. DHKW find
$N_{HI} \propto (1+\delta)^{1.43}$, whereas equation \ref{eq:NHIc} gives  
$N_{HI} \propto (1+\delta)^{1.37}$ for the value of $\alpha$ found by
DHKW ($\alpha \approx 0.5$). In the simulation of DHKW,
$N_{HI}(z=3)/N_{HI}(z=2) \approx 3.7$ and 
$N_{HI}(z=2)/N_{HI}(z=0) \approx 6.3$, while
equation \ref{eq:NHIc} yields values of 5.1 and 4.1, assuming that 
$\Gamma$ evolves according to the model of Haard \& Madau (1996),
which was used by DHKW, and a constant temperature.
In summary, equation \ref{eq:NHIc} matches the results produced by
hydrodynamical simulations very well.

\subsection{Radial sizes}
The radial sizes of absorbers can be expressed as a function of their
neutral hydrogen column densities by combining equations
\ref{eq:LJ}, \ref{eq:neutralfrac}, and \ref{eq:NHIa},
\begin{equation}
L \sim ~ 1.0\times 10^2 ~\kpc ~ 
\left ({N_{HI} \over 10^{14}~\cm^{-2}}\right )^{-1/3}
T_4^{0.41} 
\Gamma_{12}^{-1/3}
\left ({f_g \over 0.16}\right )^{2/3}.
\label{eq:L}
\end{equation}
Hence, \lya\ forest absorbers are large, $L \sim 10^2~\kpc$, with only
a weak dependence on the column density, temperature, and ionizing
background. Even clouds of $N_{HI} \sim 10^{17}~\cm^{-2}$, the highest
column density for which self-shielding can be neglected, have a
characteristic size of 10 kpc.

Equation \ref{eq:L} shows that 
the physical size of an absorber of a fixed neutral hydrogen column
density does not depend explicitly on redshift. The main time
dependence comes from the 
dependence on the ionization rate, which is thought to peak in the
range $z\sim 2$--3.  From redshift $z\sim 3$ to 0 the
ionization rate probably decreases by a factor 10--30, which
implies that the physical sizes of $z\sim 0$ absorbers are about a
factor 2 to 3 greater than those of $z\sim 3$ absorbers with the same
$N_{HI}$. The comoving sizes, however, are smaller at
$z\sim 0$ than at $z\sim 3$ by a factor 4/3 to 2.

If the absorber has not decoupled from the Hubble expansion, then the
differential Hubble flow across the absorber broadens the absorption
line, leading to a Hubble width of $b_H \sim H(z) L/2$, where $H$ is
the Hubble parameter. The Hubble width is comparable to the thermal
width, $b_T \equiv \sqrt{2kT/m} \approx 12.8\,T_4^{1/2}~\kms$, for
$N_{HI} \sim 10^{13}$--$10^{14}~\cm^{-2}$.  In reality, absorbers
are not expanding freely and peculiar velocity gradients will change
the redshift space sizes of the absorbers. Note, however, that
peculiar velocity gradients cannot decrease the width of a line 
below the thermal width. The minimum line width as a function
of the column density can therefore be used to measure the temperature
as a function of the density (Schaye et al.\ 1999; Ricotti, Gnedin, \& Shull
2000; Bryan \& Machacek 2000). Finally, it is worth noting
that the Hubble width increases almost as rapidly with temperature as
the thermal broadening width, $b_H \propto T^{0.41}$ compared to
$b_T \propto T^{0.5}$.

It is easily demonstrated that $b_H \sim b_T$ for a density contrast
$\delta \sim 0$: since $c_s \sim b_T$ and $t_{\rm dyn} \sim 1/H$,
hydrostatic equilibrium implies $b_H \sim HL \sim H c_s / t_{\rm
dyn} \sim b_T$. Similarly, it follows that for 
underdense absorbers $c_s < b_H$, i.e.\ sound waves cannot overcome
the differential Hubble expansion across the absorber. This is just
another way of showing that the fundamental assumption underlying this
work, namely that the gas clouds will typically not be far from local
hydrostatic equilibrium, breaks down for underdense absorbers.

Equations \ref{eq:NHIb} and \ref{eq:L} predict the existence of a
population of low column density ($N_{HI} \la 10^{13}~\cm^{-2}$ at
$z\sim 3$) absorption lines that arise in underdense gas and are predominantly
Hubble broadened. They should be broad and shallow and their profiles
should deviate substantially from thermal profiles. Their characteristic
sizes are greater than the sound horizon ($\Delta v \sim 10^2~\kms$),
which implies that the formalism 
developed in this paper is not well-suited to describe them.
These lines trace the underlying matter distribution and can be regarded
as a ``fluctuating Gunn-Peterson effect''. 
This means that large-scale correlations in the \lya\ absorption can
be used to constrain the primordial power spectrum (Croft et al.\
1998).


\subsection{Masses}

If the absorbers are spherical, then their characteristic mass is
$M_J \equiv \rho L^3$. Hence, the characteristic mass in gas is,
\begin{equation}
M_g \sim 8.8 \times 10^8 ~\Msun~
\left ({N_{HI} \over 10^{14}~\cm^{-2}}\right )^{-1/3}
T_4^{1.41} 
\Gamma_{12}^{-1/3}
\left ({f_g \over 0.16}\right )^{5/3},
\label{eq:mass}
\end{equation} 
and the total mass is a factor $1/f_g$ higher. Of course, the
absorbers may very well be far from spherical, in which case they can
have masses very different from equation \ref{eq:mass}. Indeed,
one would not expect mildly overdense regions, $\delta \la
10$, to be spherical, as the collapse will generally not be synchronized
along all three spatial dimensions. According to equation
\ref{eq:NHIb}, such regions will give rise to absorption lines with
column densities $N_{HI} \la 10^{15}~\cm^{-2}$ at $z\sim 3$.

\section{Cosmological implications}
\label{sec:implications}
From the observed incidence of absorption lines one can compute the
total mass in neutral hydrogen, $\Omega_{HI}$, by integrating over the
column density distribution. Similarly, if the ionization correction
is known, then the total baryon density in photoionized gas can be
computed. Furthermore, the distribution of mass as a function of
density can be derived from the differential column density
distribution. The results obtained in the previous section make it
possible to do this since equations \ref{eq:neutralfrac} and
\ref{eq:NHIa} can be combined to yield the neutral fraction as a
function of the neutral hydrogen column density.

\subsection{Contribution to $\Omega_b$}
\label{sec:omegab}
The mean gas density relative to the critical density can be obtained
from the neutral hydrogen column density distribution,
\begin{equation}
\Omega_g = {8\pi G m_H\over 3H_0 c(1-Y)}
\int N_{HI} {n_H \over n_{HI}} f(N_{HI},z) \,dN_{HI},
\label{eq:omegaba}
\end{equation}
where $f(N_{HI},z)$ is the number density of absorption lines per unit
absorption distance $X$ (a dimensionless quantity defined by equation
\ref{eq:f} below), and per unit column density. 
Observational results are usually expressed in terms of this
function. Its relation to the quantity that is actually observed, the
number density per unit redshift, $dn/dz$, depends on the assumed
cosmology: 
\begin{equation} 
f(N_{HI},z) \equiv {d^2 n \over dN_{HI}dX}
\equiv {d^2 n \over dN_{HI}dz} {H(z) \over H_0}{1 \over (1+z)^2}.
\label{eq:f}
\end{equation}
Combining equations \ref{eq:neutralfrac}, \ref{eq:NHIa} and
\ref{eq:omegaba} yields
\begin{equation}
\Omega_g \sim 2.2 \times 10^{-9} h^{-1}
\Gamma_{12}^{1/3}
\left ({f_g \over 0.16}\right )^{1/3}
T_4^{0.59} 
\int N_{HI}^{1/3} f(N_{HI},z) \,dN_{HI},
\label{eq:omegabb}
\end{equation} 
where $h \equiv H_0 / (100~\kms\,\Mpc^{-1})$.
If the gas is not isothermal, then $T$ will depend
on $N_{HI}$ and should therefore be moved inside the integral.
Note that if $f_g = \Omega_b/\Omega_m$, then equation \ref{eq:omegabb}
agrees with the scaling relation familiar from the fluctuating
Gunn-Peterson approximation, $\Omega_g \propto \Gamma^{1/2}h^{-3/2}$
(e.g., Rauch et al.\ 1997).

Observations indicate that at high redshift the column density
distribution can be fit by a power-law, $f = B\,N_{HI}^{-\beta}$ with
$\beta \approx 1.5$, over a wide range of column densities (Tytler
1987). The fact that a single power-law appears to be a reasonable
approximation is clear from Figure 1, which shows the
observed distribution at $z\approx 2.8$. Hu et al.\ (1995) find that
their data (solid points in Figure 1) is best fitted by a power-law
with $B = 5.3\times 10^7$ and $\beta = 1.46$ (for
$(\Omega_m,\Omega_\Lambda) = (0.3,0.7)$), and similar values are
reported by others (e.g., Kirkman \& Tytler 1997). This fit is shown
as the dashed line in Figure 1. Assuming
$(T_4,\Gamma_{12},f_g,h)=(2.0,1.0,0.16,0.65)$, equation
\ref{eq:omegabb} gives $\Omega_gh^2(\log N_{HI} = 13$--$17.2) \approx
0.015$. Comparing this to the value derived from high redshift
observations of the deuterium abundance, $\Omega_b h^2 \approx 0.020$
(O'Meara et al.\ 2001), we see that this leaves little room for stars,
collisionally ionized gas, and gas in absorbers with $\log N_{HI} < 13$
or $\log N_{HI} > 17.2$. Clearly, a large fraction 
of the baryons reside in the \lya\ forest.

It is important to note that it was implicitly assumed that the
measured column densities of the absorption lines are similar to the
true column densities. This assumption may, however, be incorrect for
low column density lines. Observationally, the column density
distribution is determined by decomposing absorption spectra into sets
of Voigt profiles. In high-resolution spectra with a high
signal-to-noise ratio, multiple Voigt components are usually required
to obtain a satisfactory fit for a single absorption line. Low column
density components are often placed in the wings of absorption
lines. If these spurious lines were to dominate the column density
distribution below some value of $N_{HI}$, then it would not be
possible to use that part of the
distribution to estimate the baryon density.

Another, related problem is that Monte Carlo simulations are often
used to ``correct'' the observed column density distributions for the
fraction of low column density lines that are thought to be
``missing'' because they are heavily blended with stronger lines.
This procedure can result in an unphysical extension of the power-law
distribution to arbitrary low column densities. In high quality
spectra these corrections become important below $N_{HI} \sim 10^{13}
~\cm^{-2}$ at $z\sim 3$ (e.g., Hu et al.\ 1995). The data points shown
in Figure 1 indicate the observed distribution, the ``corrected''
Hu et al.\ points would all fall on the dashed line. Unless current
estimates of the baryon density are significantly in error, the
distribution of column densities of \emph{real} absorption lines,
i.e.\ excluding Voigt components in the wings of absorption lines,
cannot extend to much lower column densities. 

Finally, it has to be kept in mind that the value obtained for
$\Omega_g$ is uncertain to a factor of order unity because of the
uncertainties in the values assumed for the various parameters and
because of the fact that there is some arbitrariness in the value of
the dimensionless coefficient in the definition
of the Jeans length. These uncertainties do, however, not affect the
derived scaling relations and the shape of the mass
distribution, which is the subject of the next section.

\subsection{Mass distribution}
It is interesting to look at the
differential distribution of the baryon density.
If $f(N_{HI},z) \propto N_{HI}^{-\beta}$, then equation
\ref{eq:omegabb} implies $\Omega_b 
\propto \int N_{HI}^{4/3-\beta}\,d\ln N_{HI}$. Hence, if $\beta <
4/3$, then the mass per decade of column density is an
increasing function of $N_{HI}$ and similarly, since $d\ln N_{HI}
\propto d\ln n_H$ (equation \ref{eq:NHIa}), the mass per decade of density
is an increasing function of $n_H$.
There is evidence for deviations from a single power-law, the column
density distribution seems to steepen somewhat 
\begin{inlinefigure}
\centerline{\resizebox{\colwidth}{!}{\includegraphics{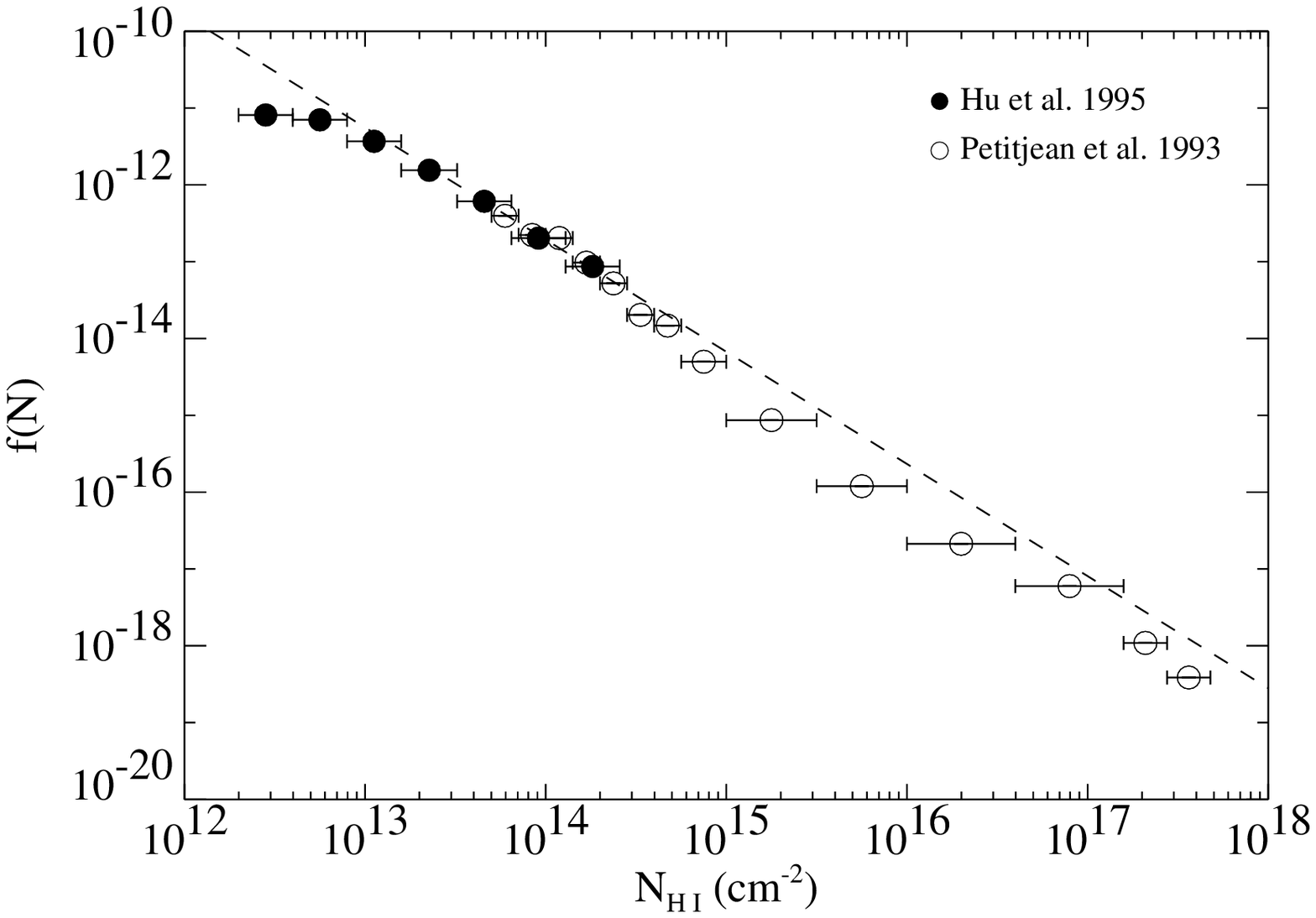}}}
\figcaption[f1.eps]{%
Observed neutral hydrogen column density
distribution. Solid points are from Hu et al.\ (1995), \emph{not}
``corrected'' for incompleteness; open circles are
taken from the compilation by Petitjean et al.\ (1993). The dashed line
is the power-law fit given by Hu et al., which is the best fit to
their ``corrected'' data in the regime $2\times 10^{12}$--$3\times
10^{14}~\cm^{-2}$, $f(N_{HI}) = 5.3 \times 10^7~N_{HI}^{-1.46}$ (the
normalization has been modified to that corresponding to a LCDM
universe, $(\Omega_m,\Omega_\Lambda) = (0.3,0.7)$). The mean redshift
of both samples is 2.8.}
\label{fig:fn}
\end{inlinefigure}
above $10^{14}~\cm^{-2}$ (Kim et al.\ 1997) before flattening to $\beta
\approx 1.32$ at $N_{HI} \ga 10^{16}~\cm^{-2}$ (Petitjean et al.\
1993). These trends are visible in Figure 1, although they are
somewhat difficult to see in a log-log plot of $f(N_{HI})$ (they would
have been more apparent if, contrary to convention, the quantity ${d^2n \over
d\log N_{HI} dX}$ had been plotted instead of ${d^2n \over d N_{HI}
dX}$). Although the reality of the deviations from the single power-law is
well established, their physical origin is not understood. In this section
it will be shown that the shape of the column density distribution
reflects the shape of the mass-weighted probability density
distribution for the gas density, i.e.\ the mass distribution of the
gas as a function of its density.

The mass distribution of the photoionized gas, $d\Omega_g/d\log(1+\delta)$,
can be derived from the observed column density distribution using
equations \ref{eq:NHIb} and \ref{eq:omegabb}. The results for the
data points shown in Figure 1 are plotted in Figure 2. The data
points were computed using the same parameter values as were used in
the previous section. Points corresponding to underdense gas ($\delta < 0$)
are less reliable because the assumption of local hydrostatic
equilibrium may break down, as was discussed before. Furthermore, the
corresponding column densities are so low ($N_{HI} <
10^{13}~\cm^{-2}$, column densities are indicated on the top axis of
Figure 2) that many lines may have been lost in
the noise. The two highest density data points are less reliable
because the effect of self-shielding is no longer negligible.
From equation \ref{eq:omegabb} it can be seen that all points are
proportional to the quantity
$h^{-1}\Gamma_{12}^{1/3}f_g^{1/3}T_4^{0.59}$. Changing 
this quantity (or adding a dimensionless coefficient in the definition
of the Jeans length) is equivalent to shifting the distribution
vertically, but does not change the shape of the distribution.

In calculating $d\Omega_g/d\log(1+\delta)$ from $f(N_{HI})$, both the gas
fraction $f_g$ and the temperature $T$ were assumed to be independent
of $N_{HI}$. Hence, the shape of the distribution shown in Figure 2 is
determined by the shape of the column density distribution. If the
latter had been a strict power-law with spectral index $\beta$, then
the former would also have been a power-law, but with index
$4/3-\beta$. Hence, the shape of the matter distribution function
traces the deviations from the single power-law in 
\begin{inlinefigure}
\centerline{\resizebox{\colwidth}{!}{\includegraphics{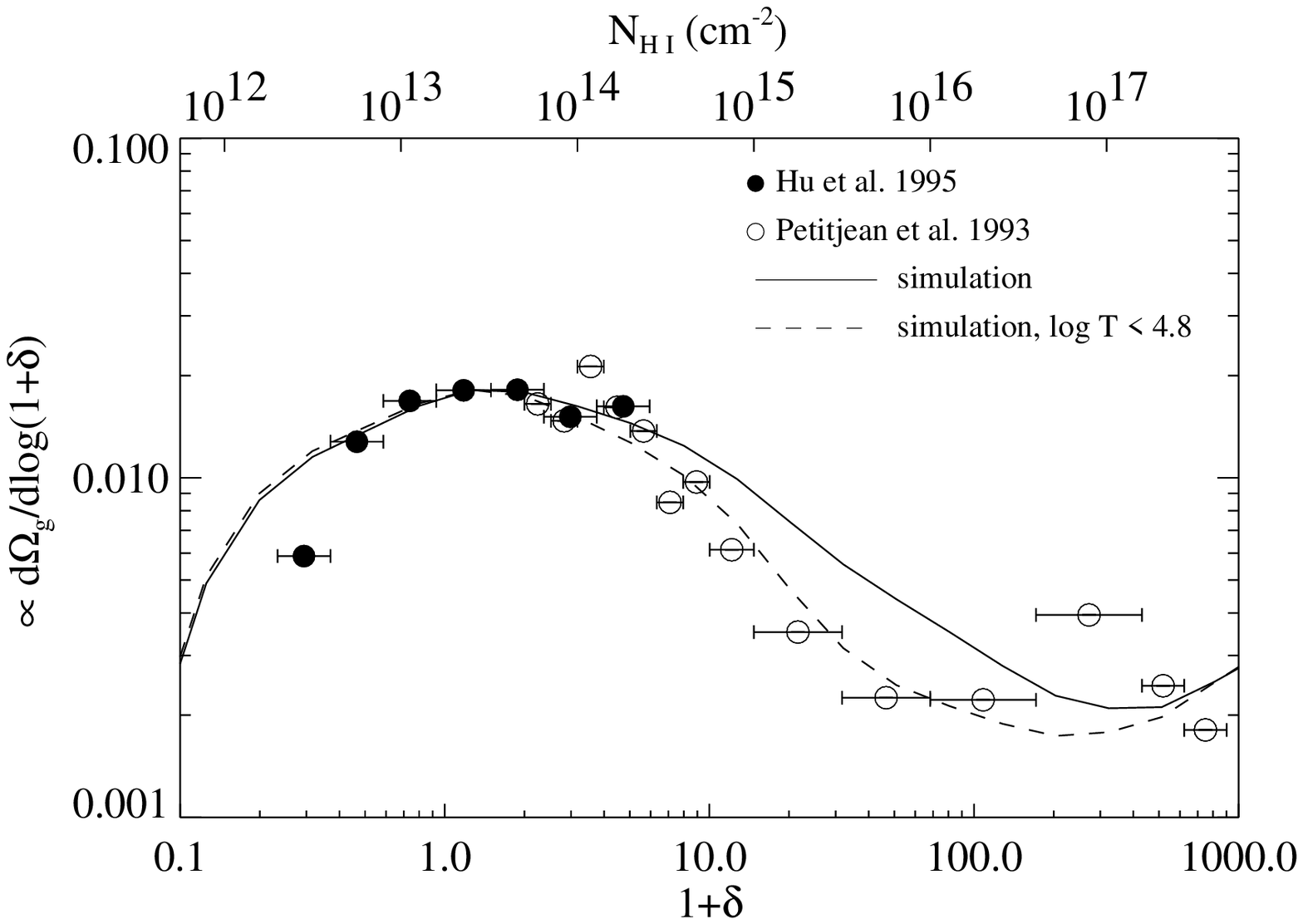}}}
\figcaption[f2.eps]{%
Mass distribution of photoionized
gas. Data points are computed from the
observed column density distribution and
equation \ref{eq:omegabb}, using the same parameter values as were
used in section \ref{sec:omegab}:
$(T_4,\Gamma_{12},f_g,h,\Omega_m,\Omega_\Lambda)=(2.0,1.0,0.16,0.65,0.3,0.7)$.
The solid line indicates the mass
distribution of all the gas in a hydrodynamical simulation at
$z=2.75$, the dashed line denotes the distribution of the gas that has
a temperature $\log T < 4.8$. The curves have been normalized to have
the same maximum as the Hu et al.\ (1995) points. The shape of the
derived mass distribution suggests that gravitational instability is
responsible for the growth of structure. The fall-off at
$\delta \sim 10$ ($\log N_{HI} \sim 14.5$) can then be attributed to the
onset of rapid, non-linear collapse and the flattening at $\delta \sim
10^2$ ($\log N_{HI} \sim 10^{16}$) to virialization. The shape of the
distribution agrees well with the shape of the gas 
distribution in the hydrodynamical simulation (solid curve), in
particular if hot gas, for which collisional ionization is important,
is excluded (dashed curve).}
\label{fig:pdf}
\end{inlinefigure}
Figure 1. The
turnover at $N_{HI} \la 10^{13}~\cm^{-2}$ may thus be real and be caused by
the turnover of the density distribution below the mean
density. However, as discussed before, both the observations and the
physical model are unreliable in this regime. The steepening at
$N_{HI} \sim 10^{14.5}~\cm^{-2}$ reflects the fall-off in the
density distribution due to the onset of rapid, non-linear collapse. Finally,
the flattening at $N_{HI} \ga 10^{16}~\cm^{-2}$ can be attributed to
the flattening of the density distribution at $\delta \ga 10^2$ due to
the virialization of collapsed matter.

To show that the derived matter distribution is consistent
with what would be expected from gravitational instability, the
distribution of gas in a hydrodynamical simulation at $z=2.75$ is
indicated by the solid line in Figure 2. The simulation, which was
kindly provided by T.~Theuns, has the same cosmological parameters
as were used to calculate the data points. It has the same
ionization background as the simulation described in Schaye et al.\
(2000), and the temperature of the low-density gas agrees with the
measurements of those authors. To compare the shape of the matter
distribution functions, the simulation data has been shifted
vertically so that it has the same maximum as the Hu et 
al.\ (1995) points. 

The shape of the simulated matter distribution
agrees fairly well with the distribution derived from the
observations, although it seems to overpredict the mass fraction in
the range $\delta \sim 10$--100. Most of this excess can in fact be
attributed to hot gas for which collisional ionization is
important. This gas is much more highly ionized than cooler,
photoionized gas of the same density and will therefore give rise to
absorption lines of much lower column densities than predicted by
equation \ref{eq:NHIb}. The dashed curve indicates the density
distribution of gas cooler than $10^{4.8}~\K$, a temperature
sufficiently low for collisional ionization to be negligible. The
agreement between this distribution and the points derived from the
column density distribution of Petitjean et al.\ (1993) is remarkable.  

The slope of the column density distribution over the range $N_{HI}
\sim 10^{13}$--$10^{14}~\cm^{-2}$ is found to steepen from $z\sim
2$ to $z\sim 0$ (e.g., Dav\'e \& Tripp 2001; Penton, Shull, \& Stocke 
2000; Kim, Cristiani, \& D'Odorico 2001; Weymann et al.\ 1998). 
If gravitational instability is
responsible for the shape of the gas density distribution, then this
distribution will be similar at $z\sim 0$ and $z\sim 3$, provided it
is expressed as a function of the density relative to the
cosmic mean. Hence, the change in the slope of the column density
distribution must mainly be due to the evolution of the function
$N_{HI}(\delta)$, which is determined by the expansion of the universe and the
evolution of the ionizing background. Equation \ref{eq:NHIb} shows
that changes in $z$ or $\Gamma$ are equivalent to shifting the
$N_{HI}$-axis (the upper x-axis) in Figure 2 relative to the
$(1+\delta)$-axis (the lower x-axis).

In section \ref{sec:densities}
it was shown that the neutral hydrogen 
column density corresponding to a fixed density contrast is about a
factor $5\times 10^2\,\Gamma(z=0)/\Gamma(z=3) \sim 5$--50 lower at
$z\sim 0$ than at $z\sim 3$. At $z\sim 3$, the distribution is found
to be steepest in the range $\log N_{HI} \sim 14.5$--15.5, which
corresponds to $\delta \sim 5$--25. This same range in density
contrast corresponds to $\log N_{HI} \sim 13.2$--14.2 at $z\sim
0$. Hence, the steepening of the slope of the column density
distribution has the same origin at low and high redshift, it
is caused by the fall-off in the probability distribution of the gas
density at the onset of rapid, non-linear collapse. Similarly, one
would expect the flattening observed at $\log N_{HI} \ga 16$ at $z\sim
3$ to shift to $\log N_{HI} \ga 15$ by $z\sim 0$, in agreement with
the observations of Weymann et al.\ (1998).

\section{Conclusions}
\label{sec:conclusions}

In the last decade semi-analytic models and numerical simulations
were used to show that the \lya\ forest is produced naturally in
cold dark matter models with initially scale invariant, adiabatic
fluctuations. Low column density absorption lines are predicted to
arise in the smoothly fluctuating, photoionized IGM, which contains a
large fraction of the baryons in the universe.
This paper demonstrates that this physical picture is in
fact generic and can be derived using straightforward physical
arguments, without making any specific assumptions about the
cosmology, the presence and distribution of dark matter
or the mechanism for structure formation. 

The key physical argument is that the sound crossing time
($t_{\rm sc}$) along any sightline intersecting a cloud, is in general
of order the local dynamical time ($t_{\rm dyn}$),
regardless of the shape of the cloud and regardless of whether the
cloud as a whole is in dynamical equilibrium. This argument
holds if the characteristic size (the scale over 
which the density is of order the maximum density) is smaller than the
sound horizon, which is
true for overdense absorbers (except for a short period after
reionization). If, for whatever reason, a large
departure from local hydrostatic equilibrium occurs, then local
equilibrium will be restored on a timescale $\min(t_{\rm sc},t_{\rm
dyn})$, which is much smaller than the Hubble time if the absorber is
overdense and out of equilibrium.
The only way to systematically violate the condition $t_{\rm sc} \sim
t_{\rm dyn}$, is to
postulate that all \lya\ forest absorbers are pressure confined, a
proposal that has been ruled out on both observational and theoretical
grounds (e.g., Rauch 1998), that they are rotationally supported
\emph{and} that their spin axes are always nearly perpendicular to the
line of sight, or that the fraction of baryons in quasi-stable ($t
\sim t_{\rm sc} \sim t_{\rm dyn} < H^{-1}$) gas clouds is negligible
compared with the fraction in transient ($t \sim \min(t_{\rm sc},
t_{\rm dyn}) \ll \max(t_{\rm sc}, t_{\rm dyn})$) clouds.
  
Using just the condition $t_{\rm sc} \sim t_{\rm dyn}$ and an
estimate of the temperature ($T \sim 10^4~\K$) and ionization rate
($\Gamma \sim 10^{-12}~\s^{-1}$ at $z\sim 3$), it was shown that
unless the ratio of baryonic to collisionless matter is orders of
magnitude lower than is generally assumed, the \lya\ forest absorbers
must be extended structures of 
low overdensity. Furthermore, the conclusion that the low column density 
absorbers contain a large fraction of the baryons, follows directly
from the observed column density distribution.

Scaling relations were derived for the radial sizes and characteristic
densities of the absorbers. It was shown that if the parameters of the
model are set to the values appropriate for the
specific model that was simulated by Dav\'e et al.\ (1999), then the density
- column density relation agrees very well with their empirical fitting
formula. Finally, the mass-weighted probability distribution of the
gas density 
was derived from the observed column density distribution. This lead
to several new insights: the shape of the column density distribution,
in particular the deviations from a single power-law, and its
evolution with redshift, reflect the shape of the matter distribution,
which agrees remarkably well with the shape of the distribution
produced by the growth of structure in an expanding universe via
gravitational instability.

This work demonstrates that Jeans smoothing is central to the
small-scale structure of the \lya\ forest. It is therefore likely that
studies in which the effect of gas pressure is neglected,
or in which the Jeans length is varied only in time, are only reliable
on scales greater than the sound horizon. On these scales ($\gg
10^2~\kms$), gas pressure is unimportant and the gas should be a good
tracer of the large-scale distribution of the matter.

It is important to note that the observational support for the current
physical picture of the \lya\ forest is much less strong for the high
column density lines ($N_{HI} \ga 10^{16}~\cm^{-2}$) than it is for
the low column density forest, which formed the main focus of this
work. It is for example possible, perhaps even likely, that galactic
winds affect some of the high-density gas that gives rise to the
strong absorption lines (e.g., Theuns, Mo, \& Schaye 2001).

Although the derived properties of the \lya\ forest absorbers are not
particularly sensitive to the presence of dark matter, dark matter may
still be required to form the observed structure in the forest from
the small-amplitude fluctuations ($\delta \sim 10^{-5}$) observed at
$z\sim 10^3$ in the cosmic microwave background. Indeed, it is
well-known that if gravitational instability is responsible for
structure formation, then dark matter (or a modification of gravity)
is required to grow the observed (non-linear) structure.
Here, the shape of the matter distribution function was derived
directly from the observed column density distribution using the
analytic model, which does not make any assumptions about the
mechanism for structure formation. The fact that the derived shape
agrees with the expectations from gravitational instability in an
expanding universe, lends credence to the use of ab-initio
hydrodynamical simulations to investigate the detailed properties of
the forest and their dependence on the parameters of the models.

\acknowledgments

I am grateful to Tom Theuns for giving me permission to use one of his
simulations for figure 2. It is a pleasure to thank Anthony Aguirre,
John Bahcall, Jordi Miralda-Escud\'e, and Eliot Quataert for
discussions and a careful reading of the manuscript. This work was
supported by a grant from the W. M. Keck Foundation.

\appendix
\section{Full equations}
In the equations appearing in the main text, numerical values were
inserted for the natural constants, as well as for many other physical
parameters. For reference, a list of the full expressions is listed
below. The contributions to the gas and electron densities of elements
other than hydrogen and helium are neglected.  

\subsection{Hydrodynamics}

The dynamical timescale $t_{\rm dyn} \equiv 1/\sqrt{G\rho}$ is
\begin{equation}
t_{\rm dyn} = \sqrt{{(1-Y)f_g \over G m_H n_H}}.
\end{equation}
The sound crossing timescale $t_{\rm sc} \equiv L/c_s$  is
\begin{equation}
t_{\rm sc} = L \sqrt{{\mu m_H \over \gamma k T}}.
\end{equation}
The condition $t_{\rm dyn} = t_{\rm sc}$ defines the Jeans length,
\begin{equation}
L_J = \left ({\gamma k \over  \mu m_H^2 G}\right )^{1/2}
(1-Y)^{1/2} f_g^{1/2} n_H^{-1/2}
T^{1/2}.
\label{eq:LJnH_app}
\end{equation}
The gaseous Jeans mass $M_{g,J} \equiv \rho_g L_J^3$ is
\begin{equation}
M_{g,J} = \left ({\gamma k \over  \mu m_H^{4/3} G}\right )^{3/2}
(1-Y)^{1/2} f_g^{3/2} n_H^{-1/2} T^{3/2},
\label{eq:MJnH_app}
\end{equation}
and the total Jeans mass is a factor $1/f_g$ greater. The Jeans column
density in hydrogen $N_{H,J} \equiv n_H L_J$ is 
\begin{equation}
N_{H,J} = \left ({\gamma k \over \mu m_H^2 G}\right )^{1/2} 
(1-Y)^{1/2} f_g^{1/2} n_H^{1/2} T^{1/2}.
\end{equation}

\subsection{Equations valid for an optically thin plasma}

The neutral fraction of a highly ionized, optically thin gas is
\begin{equation}
{n_{HI} \over n_H} = n_H {1-Y/2 \over 1-Y}
{\beta_{HII} \over \Gamma}.
\label{eq:neutralfrac_app}
\end{equation}
Substituting this into the expression for the Jeans column density
yields
\begin{equation}
N_{HI} = \left ({\gamma k \over \mu m_H^2 G}\right )^{1/2}
{1-Y/2 \over (1-Y)^{1/2}}
f_g^{1/2} \beta_{HII} \Gamma^{-1} n_H^{3/2} T^{1/2}.
\label{eq:NHInH_app}
\end{equation}
The hydrogen number density is related to the density contrast
$\delta \equiv (\rho - \bar{\rho})/\bar{\rho}$,
\begin{equation}
n_H = {3 \Omega_b H_0^2 \over 8\pi G m_H} (1-Y) (1+z)^3 (1+\delta).
\end{equation}
Using this relation, the neutral hydrogen column density can be
expressed in terms of the density contrast,
\begin{equation}
N_{HI} = \left ({3 \over 8\pi}\right )^{3/2} 
\left ({\gamma k \over \mu m_H^5 G^4}\right )^{1/2}
(1-Y)(1-Y/2) (\Omega_b H_0^2)^{3/2} (1+z)^{9/2} f_g^{1/2} \beta_{HII}
\Gamma^{-1} T^{1/2} (1+\delta)^{1.5}.
\end{equation}
Substituting equation \ref{eq:NHInH_app} into equation \ref{eq:LJnH_app}
yields an expression relating the Jeans length to the neutral hydrogen
column density,
\begin{equation} 
L_J = \left ({\gamma k \over  \mu m_H^2 G}\right )^{2/3}
\left [ (1-Y)(1-Y/2)\right ]^{1/3} f_g^{2/3} \beta_{HII}^{1/3} \Gamma^{-1/3}
T^{2/3} N_{HI}^{-1/3}.
\end{equation}
Similarly, by substituting equation \ref{eq:NHInH_app} into
equation \ref{eq:MJnH_app} the gaseous Jeans mass can be expressed in terms of
$N_{HI}$,
\begin{equation}
M_{g,J} = \left ({\gamma k \over \mu m_H^{7/5} G} \right )^{5/3}
\left [ (1-Y)(1-Y/2)\right ]^{1/3} f_g^{5/3} \beta_{HII}^{1/3} \Gamma^{-1/3}
T^{5/3} N_{HI}^{-1/3},
\end{equation}
and the total Jeans mass is again a factor $1/f_g$ greater.
The density of neutral hydrogen relative to the critical density
$\rho_c = {3H_0^2 \over 8\pi G}$, can be obtained from the number
density of absorption lines per unit absorption distance and column
density, $f(N_{HI},z)$ (see equation \ref{eq:f} for the definition
of $f$),
\begin{equation}
\Omega_{HI} = {8\pi G m_H\over 3 H_0 c} \int N_{HI} f(N_{HI},z)
\,dN_{HI}.
\end{equation}
Using equations \ref{eq:neutralfrac_app} and \ref{eq:NHInH_app}, this
can be converted into an expression for the total gas density,
\begin{equation}
\Omega_g = {8 \pi \over 3} {1 \over H_0 c}
\left ({\gamma k m_H G^2 \over \mu}\right )^{1/3}
\left [(1-Y)(1-Y/2)\right ]^{-1/3} 
f_g^{1/3} \beta_{HII}^{-1/3} \Gamma^{1/3} T^{1/3}
\int N_{HI}^{1/3} f(N_{HI},z) \,dN_{HI}.
\end{equation}
If the gas is not isothermal, then $\beta_{HII}$ and $T$ depend
on $N_{HI}$ and should be moved inside the integral.

Finally, the results derived in this paper are equally valid for
the \HeII\ \lya\ forest, provided that $N_{HI}$ is replaced by
$N_{HeII}$ using
\begin{equation}
{N_{HeII} \over N_{HI}} = {Y \over 4(1-Y)} {\beta_{HeIII} \over
\beta_{HII}} {\Gamma_{HI} \over \Gamma_{HeII}},
\end{equation}
where $\beta_{HeIII}/\beta_{HII} \approx 5.3$.

\end{document}